\documentclass[conference,a4paper]{IEEEtran}
\IEEEoverridecommandlockouts

\usepackage{cite}
\usepackage{amsmath,amssymb,amsfonts}
\usepackage{algorithmic}
\usepackage{graphicx}
\usepackage{textcomp}
\usepackage{xcolor}
\usepackage{array}
\usepackage{multirow}
\usepackage{hyperref}

\newboolean{showcomments}
\setboolean{showcomments}{false}
\ifthenelse{\boolean{showcomments}}
{ \newcommand{\mynote}[3]{
        \fbox{\bfseries\sffamily\scriptsize#1}
        {\small$\blacktriangleright$\textsf{\emph{\color{#3}{#2}}}$\blacktriangleleft$}}}
{ \newcommand{\mynote}[3]{}}

\def\BibTeX{{\rm B\kern-.05em{\sc i\kern-.025em b}\kern-.08em
    T\kern-.1667em\lower.7ex\hbox{E}\kern-.125emX}}
\begin{document}

\title{An Introductory Study on the Power Consumption Overhead of ERC-4337 Bundlers\\
}

\author{\IEEEauthorblockN{Andrei Arusoaie}
\IEEEauthorblockA{\textit{Faculty of Computer Science} \\
\textit{Alexandru Ioan Cuza University}\\
Ia\c{s}i, Romania \\
andrei.arusoaie@uaic.ro} \\
\IEEEauthorblockN{Paul-Flavian Diac}
\IEEEauthorblockA{\textit{Faculty of Computer Science} \\
\textit{Alexandru Ioan Cuza University}\\
Ia\c{s}i, Romania \\
paul.diac@info.uaic.ro}

\and
\IEEEauthorblockN{Claudiu-Nicu B\u{a}rbieru}
\IEEEauthorblockA{\textit{Faculty of Computer Science} \\
\textit{Alexandru Ioan Cuza University}\\
Ia\c{s}i, Romania \\
claudiu.barbieru@gmail.com} \\
\IEEEauthorblockN{Emanuel Onica}
\IEEEauthorblockA{\textit{Faculty of Computer Science} \\
\textit{Alexandru Ioan Cuza University}\\
Ia\c{s}i, Romania \\
emanuel.onica@uaic.ro} \\

\and
\IEEEauthorblockN{Oana-Otilia Captarencu}
\IEEEauthorblockA{\textit{Faculty of Computer Science} \\
\textit{Alexandru Ioan Cuza University}\\
Ia\c{s}i, Romania \\
oana.captarencu@info.uaic.ro} \\
\IEEEauthorblockN{Cosmin-Nicolae V\^arlan}
\IEEEauthorblockA{\textit{Faculty of Computer Science} \\
\textit{Alexandru Ioan Cuza University}\\
Ia\c{s}i, Romania \\
cosmin.varlan@info.uaic.ro}
}

\IEEEoverridecommandlockouts
\IEEEpubid{\makebox[\columnwidth]{979-8-3315-7666-0/25/\$31.00~\copyright2025 IEEE \hfill} \hspace{\columnsep}\makebox[\columnwidth]{ }}

\maketitle
\begin{abstract}

Ethereum is currently the main blockchain ecosystem providing decentralised trust guarantees for applications ranging from finance to e-government.
A common criticism of blockchain networks has 
been their energy consumption and operational costs. The switch from Proof-of-Work (PoW) protocol to Proof-of-Stake (PoS) protocol has significantly reduced this issue, though concerns remain, especially with network expansions via additional layers.
The ERC-4337 standard is a recent proposal that facilitates end-user access to Ethereum-backed applications. 
It introduces a middleware called a bundler, operated as a third-party service, where part 
of its operational cost is represented by its power consumption. 
While bundlers have served over 500 million requests in the past two years, fewer than 15 official bundler providers exist, compared to over 100 regular Ethereum access providers.
In this paper, we provide a first look at the active power consumption overhead that a bundler would add to an Ethereum access service. 
Using SmartWatts, a monitoring system leveraging Running Average Power Limit (RAPL) hardware interfaces, we empirically determine correlations between the bundler workload and its active power consumption.
\end{abstract}


\section{Introduction}
\label{sec:introduction}

Blockchain networks have significantly changed the landscape of distributed systems over the last decade by providing an accessible, decentralized trust solution. 
Users send transactions to the blockchain nodes, which maintain a public registry, referred to as \emph{the ledger}.
This ledger consists of securely chained blocks that are composed of the users' transactions. 
Blocks are proposed for addition to the ledger by blockchain nodes chosen according to a consensus protocol.
The ledger represents an immutable source of decentralized trust - confirmed transactions can be queried and verified by any network client. 
Bitcoin~\cite{nakamoto2008bitcoin} is the first blockchain implementation, used exclusively for cryptocurrency transactions.

The advent of blockchain networks has been shadowed by persistent concerns over the substantial energy requirements inherent in their initial consensus mechanisms.
Proof-of-Work (PoW) used in Bitcoin, requires nodes to compete for the block proposer role by solving a computationally difficult puzzle. 
This results in an energy consumption reported as exceeding 190 TWh per year in September 2025~\cite{bitcoinenergy}, comparable to the entire state of Thailand. 
For this reason, multiple blockchains changed their consensus algorithm to Proof-of-Stake (PoS), where the block proposer is elected using a complex protocol based on staked economic value.
Ethereum~\cite{ethereum} switched to PoS in 2022, reducing the entire network energy consumption to a reported 0.0026 TWh per year~\cite{ethereumenergy}.

Ethereum produced another major shift in the typical use of blockchain networks, providing support for generic transactions besides cryptocurrency exchanges. 
Nodes run a core component, the Ethereum Virtual Machine (EVM), capable of executing~\emph{smart contracts}, small programs translated from a high-level language to a low-level bytecode. 
Variables defined in smart contracts are part of the blockchain state, and functions changing these are triggered by transactions. 
This enables the possibility of using Ethereum as a backend support for a variety of decentralised applications (DApps). 
Examples span from financial exchanges to e-voting scenarios~\cite{DApps}.

A prominent use case are transferable tokens, which can represent assets of value, e.g., application specific currencies (fungible) or unique collectible items (non-fungible).
The widest spread token is defined by the ERC-20 standard~\cite{ERC20}, with more than 1.7M implementations~\cite{tokenstotal}.
An ERC-20 token specific to an application can be used both as currency within an application (e.g., a fidelity token for an online shop), and also traded similar to stocks, via exchange services. 
According to~\cite{somin2025cryptoasset}, close to 2 billion ERC-20 trades were executed from November 2nd, 2015 to December 31st, 2024.

Despite the above numbers, onboarding users is still a cumbersome problem to solve~\cite{Chandra22}.
Transactions in Ethereum have a cost, which depends on their computation by the EVM 
and the required storage. 
This cost is expressed in units of~\emph{gas}, which is priced dynamically in the native Ethereum currency (ETH), depending on the transactions frequency at a certain moment.
This prevents Denial-of-Service (DoS) attacks by the invocation of malicious contract code, and balances the network use.  
This essentially requires users to pay for transactions initiated by an application, which can be inconvenient, especially considering that the fee for the same transaction can vary over time. 
Moreover, some users might not possess native ETH, but only tokens specific to an application.

A solution proposed in a recent Ethereum standard is ERC-4337: Account Abstraction~\cite{ERC4337}. 
This associates with a user a Smart Contract Account (SCA), which has two primary roles: to provide a user with custom verification logic for an initiated transaction and to add flexibility to transaction payment. 
The latter allows either a conversion of the transaction fee to ERC-20 tokens or having it sponsored by a third party. 
To take advantage of the ERC-4337 functionality, communication must follow a distinct path rather than through an Ethereum node.
Transactions will be submitted to an ERC-4337 bundler middleware, as we further describe in Section~\ref{sec:background}.

The bundler performs several tasks, such as pre-validation and grouping of multiple transactions for efficiency, before sending them to the Ethereum network. 
The bundler is normally offered as a public service, gaining profit by increasing the transaction fee for its role in the communication path.  
However, there are currently less than 15 official reported ERC-4337 bundler providers~\cite{bundlers}.
We note that some of these providers, such as Alchemy or ThirdWeb, also act as regular access providers for Ethereum, where their total number exceeds 130~\cite{CompareNodes}.
A regular Ethereum access provider essentially offers access to the network via a JSON-RPC protocol on its own Ethereum nodes, charging an access fee.

\subsubsection*{Motivation}
A bundler middleware will forward the transactions to the Ethereum network. 
This requires the bundler to maintain a low latency for quality of service. 
This latency can be reduced to a minimum if the bundler middleware is hosted together with an Ethereum node.
This also removes the cost of having a separate bundler dedicated node. 
Then the bundler's operational cost is split between some variable costs specific to the protocol, i.e., advance coverage of the transaction fees, the communication overhead from received transactions, and the power consumption overhead of the bundler implementation. 

Studies on transaction costs associated with ERC-4337 (i.e., gas consumption) already exist~\cite{Lin2024}.
Handling the communication overhead is a matter of scaling common to a regular Ethereum access provider's case, e.g., it can be solved by integrating a load balancing service~\cite{Wang2022}. 
In this paper, we focus on bundler power consumption, which, to our best knowledge, has not been investigated before.
This can impact the bundler's profitability or fees for the user.
In particular, it is relevant to investigate which part of the bundlers' activity would increase power consumption, to understand whether the bundler service can be co-hosted with a regular Ethereum node.
To determine this at process granularity, we leverage SmartWatts~\cite{Fieni2020SmartWatts}, a dedicated tool. 

\subsubsection*{Paper Organisation}
In Section~\ref{sec:background} we detail the ERC-4337 flow and provide some background on the instrumentation used for measuring power consumption.
We continue in Section~\ref{sec:setup} with the description of the experimental pipeline setup for testing a bundler middleware.
In Section~\ref{sec:evaluation} we present the obtained results, and we conclude in Section~\ref{sec:conclusion}.

\section{Background}
\label{sec:background}

\begin{figure*}[ht]
  \centering
  \includegraphics[width=\textwidth]{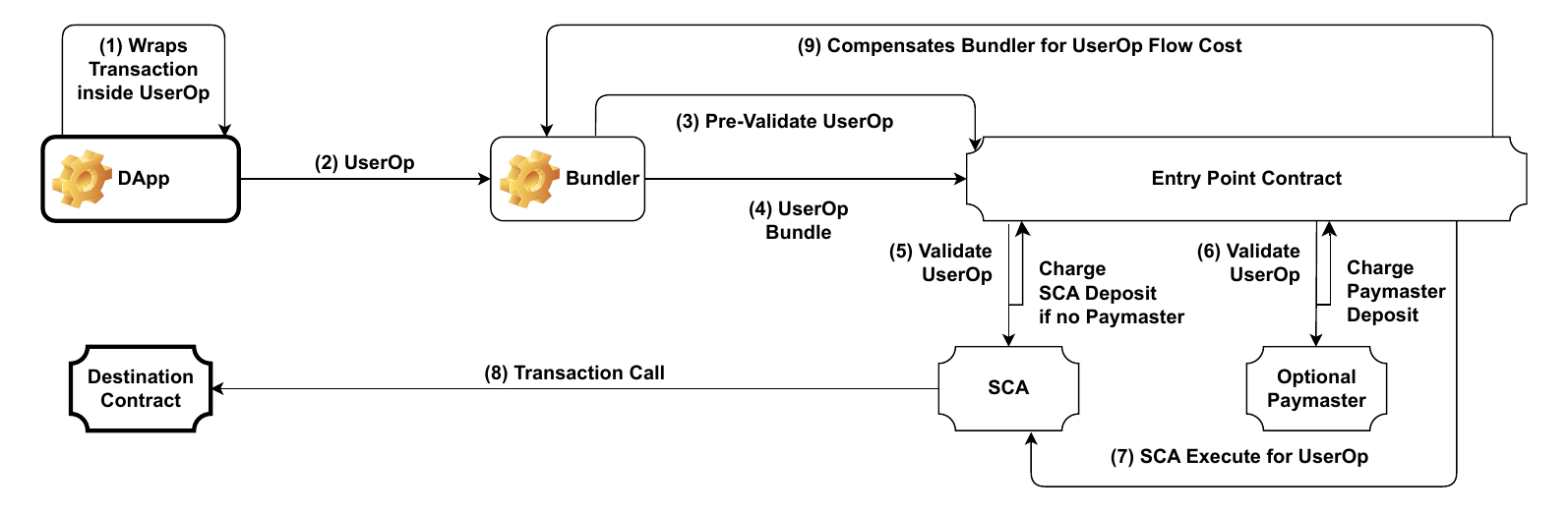}
  \caption{\label{figuserop}Transaction communication path in ERC-4337}
\end{figure*}

In this section, we first discuss the integration of bundler middleware into the communication path specific to ERC-4337. 
Then, we provide details about SmartWatts, the software instrument we use to estimate active power consumption.

\subsection{The ERC-4337 transaction path}

The communication path using an ERC-4337 bundler is displayed in Fig.~\ref{figuserop}.
First, this requires an application to wrap the transaction into a User Operation (UserOp) message format that is sent to the bundler (1,2), maintaining its intended final contract destination in the wrapped part.
The bundler performs a pre-validation of the UserOps (3), and then groups multiple such messages in a bundle forwarded towards an Entry Point contract in the Ethereum network (4).
The bundler must pay in advance the fee for this bundle transaction.
We note that some bundlers~\cite{altosafemode} allow disabling some pre-validation parts for wider compatibility with EVM nodes.

Further, the EntryPoint contract initiates a processing flow for each of the bundled UserOps.
It first performs a validation with the SCA associated with the UserOp initiator (5). 
The UserOp can optionally specify a third-party paymaster to sponsor the fee of the transaction.
Depending on this, the UserOp is also validated with the paymaster (6), and the EntryPoint charges either the SCA or the paymaster for the transaction's costs.
Then, the EntryPoint triggers a generic execution function in the SCA (7), passing the wrapped transaction data in the UserOp, which results in a regular transaction call to the destination contract (8). 
Finally, the EntryPoint compensates the bundler for the pre-funding (9).

We note that step (2) is simplified in the presented figure, some bundlers sharing a common mempool of UserOps. 
Depending on this, additional replication of the UserOp could take place, which could increase the overhead of the bundler process. 
However, currently only a few of the official bundlers support this~\cite{bundlersoverview}, and the UserOp distribution is performed via gossiping, not by broadcast, involving communication to a small number of peer nodes.  
Steps (3) and (4) are executed via remote procedure calls (RPC) using the JSON-RPC protocol exposed by the EVM running on the blockchain nodes.
Therefore, the strict overhead of the bundler process for these calls is simply the client-side RPC stub.
We observe that a consistent part of the flow's steps (5-8) does not involve the bundler node, but just the blockchain nodes handling the execution of the contracts.
Finally, step (9) is a payment transaction, also executed on the blockchain nodes. 

Intuitively, we can observe that the overhead brought by the bundler process should be limited to processing the UserOps. 
Of course, indirectly, there is a more consistent overhead added by the ERC-4337 flow to the blockchain node that runs the EVM executing the EntryPoint transactions. 
However, this overhead would exist nonetheless, irrespective of the placement of the bundler node.

\subsection{Using SmartWatts for power consumption estimation}

Power consumption of a computer is heavily influenced by the software running on the respective node~\cite{Colmant2018,Acar2016,Verdecchia2017}. 
In a general computing context, the CPU is widely recognized as a main hardware component accounting for the variations in power drawn, which correlates to the entire system consumption~\cite{Colmant2018,Rivoire2008HighLevel,Mahesri2005}. 
Systems specialised for heavy computations using dedicated hardware, such as GPUs or ASICs, also exhibit consistent power consumption. 
However, this is not the case with an ERC-4337 bundler. 

While a wide range of physical smart meter devices can be used to monitor the power drawn by a node, isolating the power consumption overhead added by a particular software is more complex.
A rough estimate can be provided by simply averaging the difference between repeated executions with or without the considered software. 
However, this does not isolate the effect of multiple factors that might impact active power consumption: variable background noise from the software dependencies or other processes, hardware events covered by CPU Dynamic Voltage/Frequency Scaling (DVFS) that enables 
CPU clock and voltage adjustment upon need, cache misses caused by the execution context, or entering C-states (low-power) when idling.
This prompted the research of various power models ~\cite{Colmant2018,Acar2016,Do2009pTop}, most of which track various system measurements and employ a machine learning approach for an accurate estimation of CPU consumption. 

In our study, we rely on SmartWatts~\cite{Fieni2020SmartWatts}, which is part of the PowerAPI toolkit~\cite{powerapiorg,Colmant2018}.
This leverages Running Average Power Limit (RAPL)~\cite{IntelSDM}, an interface for reporting the active power consumption of several hardware domains, which was introduced on Intel processors starting with the Sandy Bridge architecture and later by AMD from the Zen architecture.
The domains available on Intel processors for RAPL measurements cover power consumed by the entire CPU package, including the cores, cache levels, and any additional support included with the CPU die, such as integrated graphics. 
While the RAPL interface facilitates a software measurement of CPU power consumption, this still represents a global rough estimate that can lack accuracy~\cite{Colmant2015}.

To improve precision, SmartWatts uses additional hardware performance counters (HwPC) from the CPU’s performance monitoring units (PMU). 
Key metrics include the APERF/MPERF ratio, which compares actual to nominal core frequency and reveals turbo boost or power saving behavior; unhalted core cycles; last-level cache (LLC) misses requiring main memory access; and instructions retired (completed).

The RAPL measurements and the rest of the HwPC data are captured in the context of process separation provided on Linux kernels via the control groups (cgroups) facility, maintaining a per-group mapping. 
The monitoring first considers a static power isolation phase, which aims to determine a quiescent state corresponding to the power drawn during the system's operation at rest. 
This is done by separating measurements when either a process control activity or global idle CPU cycles exceed a certain threshold. 
The effective power model estimates global and per-group active power consumption, feeding the collected data into a Ridge regression formula.
The HwPC events are elected according to a Pearson correlation with the RAPL values that exceed a baseline value. 
The model performs dynamic recalibration whenever it is triggered by a quantified error estimation.
SmartWatts reports an accuracy of less than 3.5\% of error compared to 5\% for global RAPL~\cite{Khan2016}, and, in addition, it provides a split estimation per process group. 
\section{Bundler Pipeline Setup}
\label{sec:setup}

In this section, we describe our software setup, assembled to estimate a bundler's active power consumption overhead.
This is separated into a functional part that ensures the transaction flow from the source to execution in the destination contract and a monitoring part dealing with measurements.

\subsection{Transaction Flow Pipeline}

For the convenient support of contract deployment and transaction execution, we consider Anvil~\cite{anvil}, one of the existing Ethereum development nodes.
While this essentially provides a simulation of an actual node, it implements the main functionalities found on such a machine that deal with transaction processing, including a fully compatible EVM and an implementation of the JSON-RPC communication interface. 
It mainly abstracts away the consensus logic and various advanced components, such as optimisations for high traffic loads or specific security measures. 
Anvil can be configured to simulate block production at regular intervals, and is an appropriate choice for our purpose of measuring the bundler power drain overhead, as it exhibits a lower energy footprint than a full-fledged Ethereum node.
This permits us to establish a more stable baseline, with fewer variations in power consumption due to the simpler implementation.

For the deployed contracts required by the ERC-4337 environment, we use the first production version of the EntryPoint (v0.6), which remains the most popular on the Ethereum network~\cite{entrypoint6,entrypoint7} at the time of writing, and the corresponding reference SCA implementation (SimpleAccount)~\cite{infinitism}.
As a target destination for the transactions in our tests, we opt for a simple ERC-20 token contract~\cite{ERC20}.
We use token transfer transactions between the deployed SCAs as our workload, a widespread use case in Ethereum. 

We selected Alto~\cite{altobundler} for our evaluation, since it is one of the top-ranked bundlers in terms of processed UserOps~\cite{bundlers} and is implemented in TypeScript~\cite{bundlersoverview}. 
A ranking of programming languages energy efficiency~\cite{Pereira2017} based on generic benchmarks shows that TypeScript consumes about 21.5× more power than C (the baseline), and significantly more than Rust (1.03×), Java (1.98×), or Go (3.23×), languages used by some other bundlers. 
Although implementation details affect efficiency, a TypeScript-based bundler can be regarded at the upper bound for power consumption relative to equivalent implementations in other languages.
We assemble Alto and the above components in a pipeline resembling the transaction flow depicted in Figure~\ref{figuserop}, filling the DApp role with a simple script that sends the workload of transactions to the bundler.

\subsection{Monitoring Setup}

For monitoring the power drain, we use the RAPL and HwPC sensor and the SmartWatts estimation formula that are part of the PowerAPI toolkit~\cite{powerapiorg}.
We isolate the Anvil and Alto processes by running these in separate transient scope units. 
These function as temporary, distinct cgroups that wrap each of the two processes while they are executed, allowing the PowerAPI components to individually estimate their consumption. 
In the following section, we provide details about the configuration parameters, executed tests and obtained results.

\section{Evaluation}
\label{sec:evaluation}

We performed our experimental evaluation on a system running a lightweight version of Ubuntu 22.04.3 LTS, with Linux 6.8.0-84-generic kernel, equipped with an 8-core (16 threads) Intel i7-10870H CPU, 45W TDP, running at a baseline frequency of 2.20GHz, with 16 GB of DDR4 memory clocked at 2667 MHz, and using a 512 Gb NVMe SSD. 

We used the setup described in Section~\ref{sec:setup}, with Anvil v1.3.5 as a chain node simulator and the latest release of the Alto bundler. 
As a workload, we deployed 100 SCAs and considered bursts of ERC-20 token transfer UserOps addressed to different accounts, i.e., which can be processed in parallel. 
We fed the UserOps to the bundler using the localhost interface to abstract the system's network I/O overhead. 
Our main interest is specifically in the isolated power drain of the bundler process. 
This particularly impacts the active CPU power consumption, which we estimate using SmartWatts. 

We also used SmartWatts to measure the overhead inflicted on the chain simulator process, as well as the global CPU active power consumption.
Concerning the latter, we also report the raw RAPL CPU power measurements extracted by the PowerAPI toolkit.
We performed several tests, as further described, with UserOps burst loads duration spanning between 30 minutes to 2.5 hours.
For each, we used a sampling interval of 500ms for collecting the RAPL measurements, also used by the SmartWatts estimation.   
In addition, we also obtained a measurement of the entire system's power consumption through a physical smart meter (TP-Link Tapo P115). 
We note that this provides information at fixed coarse-grained intervals of 5 minutes, which we considered just as a rough verification measure for the software-reported CPU consumption and the correlation with the overheads increase.

\begin{table}[t]
  \caption{Weighted Average of Watts Consumption in Different Tests}
  \label{tab:pwc}
\renewcommand{\arraystretch}{1.3}
\begin{tabular}{| m{0.23\linewidth} || m{0.11\linewidth} | m{0.09\linewidth} | m{0.09\linewidth} | m{0.08\linewidth} | m{0.09\linewidth} |} 
 \hline
 Test Variations & Alto (bundler) & Anvil (chain) & Global CPU & RAPL CPU & System Total\\ 
 \hline\hline
System at Rest & 0.10 & 1.30 & 3.00  & 3.62 & 18.22 \\\hline\hline
\multicolumn{6}{|c|}{Throttling Variation with 100 UserOps Burst Load and Blocks at 15s} \\\hline
25ms Throttling & 3.22 & 5.42 & 12.85 & 13.54 & 22.54 \\\hline
50ms Throttling & 1.50 & 1.84 & 5.66 & 7.61 & 18.81 \\\hline
100ms Throttling & 0.99 & 1.92 & 4.90 & 5.17 & 18.46 \\\hline\hline
\multicolumn{6}{|c|}{UserOps Burst Load Variation with 25 ms Throttling and Blocks at 15s} \\\hline
75 UserOp Burst & 2.96 & 3.38 & 9.58	& 10.52	& 24.22 \\\hline
50 UserOp Burst & 1.54 &	2.74	& 6.24	& 6.91	& 21.88 \\\hline
25 UserOp Burst & 1.01 & 2.46	& 5.37	& 5.67	& 19.55 \\\hline\hline
\multicolumn{6}{|c|}{Block Frequency Variation with 100 UserOps Burst and 25 ms Throttling} \\\hline
5s Block & 4.68	& 5.51 & 	12.61 &	13.39	& 24.20 \\\hline
10s Block & 4.49 &	4.18 &	10.76	& 11.37 &	24.71 \\\hline
\end{tabular}
\vspace{-10pt}
\end{table}

\begin{figure*}[ht]
  \centering
  \includegraphics[width=\textwidth]{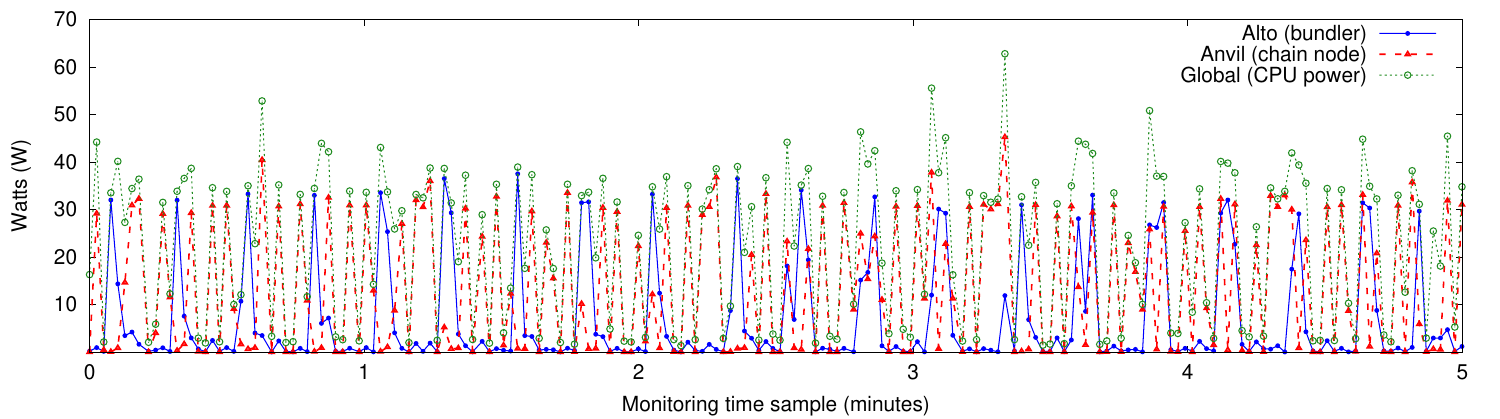}
  \caption{\label{fig:components}Power Consumption Sampling of Bundler, Chain and Global CPU (block frequency at 15s, 100 UserOps bursts, 25ms throttling per UserOp)}
\vspace{-5pt}
\end{figure*}

\begin{figure*}[ht]
  \centering
  \includegraphics[width=\textwidth]{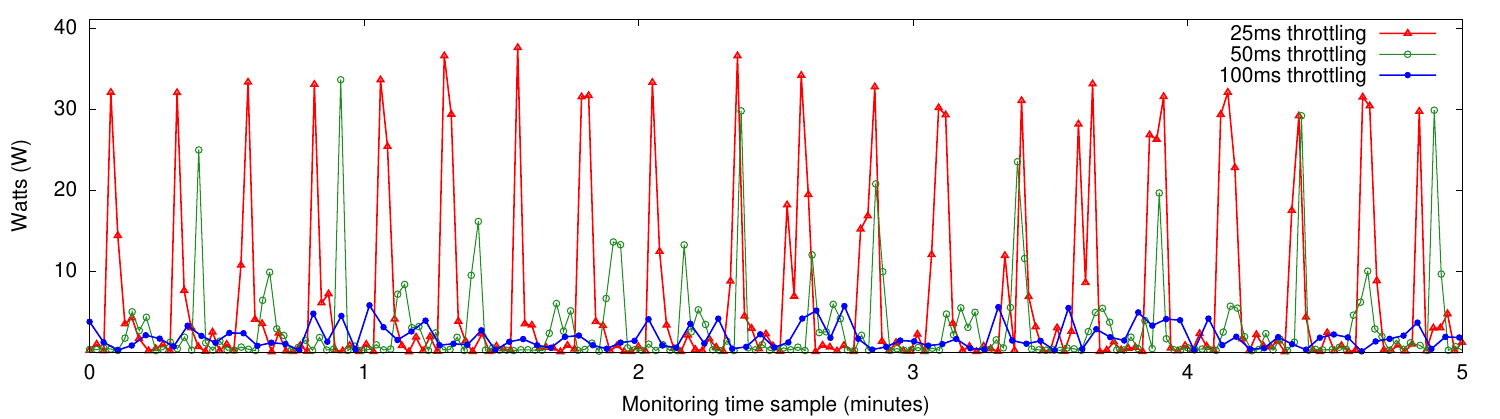}
  \caption{\label{fig:throttling}Power Consumption Sampling of Bundler at Different UserOp Throttling Rates (block frequency at 15s, 100 UserOps bursts)}
\vspace{-5pt}
\end{figure*}

\begin{figure*}[ht]
  \centering
  \includegraphics[width=\textwidth]{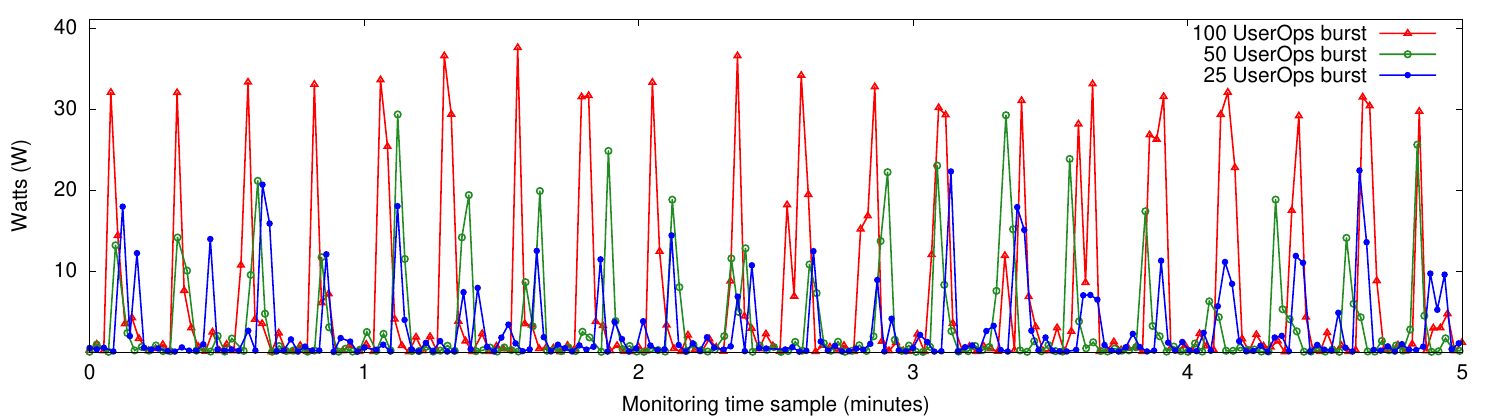}
  \caption{\label{fig:userops}Power Consumption Sampling of Bundler at Different UserOp burst loads (block frequency at 15s, 25ms throttling per UserOp)}
\vspace{-7pt}
\end{figure*}

The results of our evaluation are summarised in Table~\ref{tab:pwc}, which provides the time-scaled weighted average in Watts observed in our experiments. 
First, we measured the power consumption of the system at rest. 
This covers the idle bundler process, the chain node simulator and the rest of the background processes. 
We set Anvil to produce blocks every 15s, which occurs even when UserOps traffic and resulting transactions are absent, i.e., empty blocks.
The Alto bundler consumption is close to 0, while Anvil shows a minor 1.3W, with the whole CPU package below 4W.

In the first experiment set, we used bursts of 100 UserOps, issued in rounds triggered by block confirmation, i.e., up to 15 seconds when the entire burst set is confirmed in the current block.
Since transmission occurs locally and bursts are likely to impact the per-socket buffers, even if they do not reach the physical network interface, we introduce some throttling when sending each UserOp.
We observe that this sending rate has an impact on CPU power consumption. 
This can be noticed in Fig.~\ref{fig:components}, which presents an extracted sample from a test that uses only a small throttling value of 25ms per UserOp.

A low throttling rate permits all UserOps to be confirmed fast in the next block.
However, when sending a new burst, this causes a consistent spike in the CPU active power, reaching close to 40W for the bundler alone, although the weighted average remains under 4W for the entire test, due to the longer periods at rest.
The impact can also be observed on the Anvil node, which typically receives 3-4 bundle groups in transactions from the bundler that correlate with its respective spikes observed in the sample figure.
The CPU power drain decreases when we increase the throttling, as we observe in Fig.~\ref{fig:throttling}.
The weighted average bundler consumption drops around 1Wh when adding 100ms delay for each UserOp.
This delay has low impact on the UserOp confirmation, which still happens within a span of 2 consecutive blocks.

In Fig.~\ref{fig:userops}, we present three samples from other test series, where we vary the burst size, keeping the low throttling of 25ms.
As expected, the CPU active power decreases with the size of the burst, with a weighted average of around 1Wh for 25 UserOps, which confirms the same optimization ratio in the previous test, i.e., one fourth of the full load, similar to a four times increase in the throttling interval. 
On the other hand, we note that the overhead in power consumption is not linear with either the throttling rate or the burst load, resulting in a typical lower increase for all the monitored components, showing that the bundler scales. 
We also observe this in the final test series, as shown in Table~\ref{tab:pwc}, where we vary the block frequency to 10s and 5s. 
Although the consumption exceeds 4W due to the consequent higher burst rounds frequency, this remains sublinear to the effective increase in load. 

\section{Conclusion}
\label{sec:conclusion}

In this study, we provided a first investigation of an ERC-4337 bundler active power consumption, an important component in the Ethereum ecosystem, which can facilitate user access to the network.
From our experiments, we determined that the power consumption mainly correlates with the load of received user requests that need processing. 
While this can add an overhead to a node that runs the bundler, we observed that this overhead remains sublinear to the received load. 
The question of how profitable it would be to run a bundler node together with a regular block-building node requires further investigation of also other economic factors.
However, we believe that our research provides at least some indication helpful to approach this matter from the power consumption perspective.
We are aware that our findings in this introductory study are based on empirical observations, limited only to a single bundler implementation, although a top-tier one in terms of current use.
In future work, we intend to investigate whether other bundlers exhibit lower consumption, and perform a comparison also using different hardware support.

\section*{Acknowledgment}
This work was supported by a grant from the Romanian Ministry of Research, Innovation and Digitization, CNCS/CCCDI - UEFISCDI, project number 86/2025 ERANET-CHISTERA-IV-SCEAL, within PNCDI IV.

{
\bibliographystyle{IEEEtran}
\bibliography{biblio}

@book{DApps,
    author = {Arshdeep Bahga and Vijay Madisetti},
    title = {Blockchain Applications - A Hands-On Approach},
    year = {2017},
    publisher = {VPT}
}

@online{Chandra22,
	title = {Web3: {O}nboarding the next billion users - {T}he road ahead},
	url = {https://cointelegraph.com/news/web3-onboarding-the-next-billion-users-the-road-ahead},
	author = {Chandra, Sharat and Aggarwal, Shiv},
	year = {2022},
    lastaccessed = {September 19, 2025}
}

@online{ERC4337,
  author =       {Vitalik Buterin and Yoav Weiss and Dror Tirosh and Shahaf Nacson and Alex Forshtat and Kristof Gazso and Tjaden Hess},
  year =         {2021},
  title =        {{ERC}-4337: {A}ccount Abstraction Using Alt Mempool},
  url =          {https://eips.ethereum.org/EIPS/eip-4337},
  lastaccessed = {September 19, 2025}
}

@online{CompareNodes,
	title = {Top 139 {E}thereum {RPC} Nodes and Data {API}s},
	url = {https://comparenodes.com/protocols/ethereum/},
	year = {2025},
    lastaccessed = {September 19, 2025}
}

@misc{ethereum,
     title={Ethereum: A secure decentralised generalised transaction ledger},
     author={Wood, Gavin},
     journal={Ethereum project yellow paper},
     howpublished = {\url{https://ethereum.github.io/yellowpaper/paper.pdf}},
     lastaccessed = {September 19, 2025}
 }

@online{tokenstotal,
    title = {{ERC}-20 tracker},
    year = {2025},
    url = {https://etherscan.io/tokens},
    lastaccessed = {September 19, 2025}
}

@online{ERC20,
  author =       {Fabian Vogelsteller and Vitalik Buterin},  
  title =        {{ERC-20}: Token Standard},
  year =         {2015},
  url =          {https://eips.ethereum.org/EIPS/eip-20},
  lastaccessed = {September 19, 2025}
}

@misc{nakamoto2008bitcoin,
  title        = {Bitcoin: A Peer-to-Peer Electronic Cash System},
  author       = {Nakamoto, Satoshi},
  year         = {2008},
  howpublished = {\url{https://bitcoin.org/bitcoin.pdf}},
  lastaccessed = {September 19, 2025}
}

@online{bitcoinenergy,
    title = {Bitcoin Energy Consumption Index},
    year = {2025},
    url = {https://digiconomist.net/bitcoin-energy-consumption},
    lastaccessed = {September 19, 2025}
}

@online{ethereumenergy,
    title = {Ethereum's Energy Expenditure},
    year = {2025},
    url = {https://ethereum.org/energy-consumption/},
    lastaccessed = {September 19, 2025}
}

@article{somin2025cryptoasset,
  title  = {Crypto-asset trading on top of {E}thereum Blockchain: comprehensive dataset},
  author = {Somin, Shahar and Altshuler, Yaniv and Pentland, Alex},
  journal= {Scientific Data},
  year   = {2025},
  volume = {12},
  pages  = {1407},
  doi    = {10.1038/s41597-025-05662-w}, 
  lastaccessed = {September 19, 2025}
}

@online{bundlers,
    title = {{ERC}-4337 Bundlers - {E}thereum Network},
    year = {2025},
    url = {https://www.bundlebear.com/erc4337-bundlers/ethereum},
    lastaccessed = {September 19, 2025}
}

@article{Colmant2018,
title = {The next 700 {CPU} power models},
journal = {Journal of Systems and Software},
volume = {144},
pages = {382-396},
year = {2018},
issn = {0164-1212},
doi = {https://doi.org/10.1016/j.jss.2018.07.001},
author = {Maxime Colmant and Romain Rouvoy and Mascha Kurpicz and Anita Sobe and Pascal Felber and Lionel Seinturier},
}

@article{Acar2016,
  TITLE = {{The Impact of Source Code in Software on Power Consumption}},
  AUTHOR = {Acar, Hayri and Alptekin, G{\"u}lfem I and Gelas, Jean-Patrick and Ghodous, Parisa},
  JOURNAL = {{International Journal of Electronic Business Management}},
  PUBLISHER = {{Electronic Business Management Society, Taiwan}},
  VOLUME = {14},
  PAGES = {42-52},
  YEAR = {2016}
}

@inproceedings{Verdecchia2017,
author = {Verdecchia, Roberto and Procaccianti, Giuseppe and Malavolta, Ivano and Lago, Patricia and Koedijk, Joost},
title = {Estimating energy impact of software releases and deployment strategies: the {KPMG} case study},
year = {2017},
isbn = {9781509040391},
publisher = {IEEE Press},
booktitle = {Proceedings of the 11th ACM/IEEE International Symposium on Empirical Software Engineering and Measurement},
pages = {257–266},
}

@misc{IntelSDM,
 author = {Intel},
 month = {December},
 title = {Intel 64 and {IA-32} Architectures Software Developer’s Manual},
 volume = {3},
 year = {2022}
}

@inproceedings{Colmant2015,
author = {Colmant, Maxime and Kurpicz, Mascha and Felber, Pascal and Huertas, Lo\"{\i}c and Rouvoy, Romain and Sobe, Anita},
title = {Process-level power estimation in {VM}-based systems},
year = {2015},
isbn = {9781450332385},
publisher = {Association for Computing Machinery},
booktitle = {Proceedings of the Tenth European Conference on Computer Systems},
articleno = {14},
numpages = {14},
location = {Bordeaux, France},
}

@inproceedings{Rivoire2008HighLevel,
  author    = {Suzanne Rivoire and Parthasarathy Ranganathan and Christos Kozyrakis},
  title     = {A Comparison of High-Level Full-System Power Models},
  booktitle = {Proceedings of the 2008 Workshop on Power Aware Computing and Systems},
  year      = {2008},
  publisher = {USENIX Association},
}

@InProceedings{Mahesri2005,
author="Mahesri, Aqeel
and Vardhan, Vibhore",
title="Power Consumption Breakdown on a Modern Laptop",
booktitle="Power-Aware Computer Systems",
year="2005",
publisher="Springer Berlin Heidelberg",
pages="165-180",
isbn="978-3-540-31485-1"
}

@inproceedings{Do2009pTop,
  author       = {Thanh Do and Suhib Rawshdeh and Weisong Shi},
  title        = {{pTop}: A Process‐level Power Profiling Tool},
  booktitle    = {Proceedings of the 2nd Workshop on Power Aware Computing and Systems},
  year         = {2009},
  publisher    = {USENIX Association},
}

@inproceedings{Fieni2020SmartWatts,
  author       = {Guillaume Fieni and Romain Rouvoy and Lionel Seinturier},
  title        = {SmartWatts: Self‐Calibrating Software‐Defined Power Meter for Containers},
  booktitle    = {Proceedings of the 20th IEEE/ACM International Symposium on Cluster, Cloud and Internet Computing},
  year         = {2020},
  month        = {May},
  publisher    = {IEEE and ACM},
  pages        = {45--54},
}

@article{Khan2016,
author = {Khan, Kashif Nizam and Ou, Zhonghong and Hirki, Mikael and Nurminen, Jukka K. and Niemi, Tapio},
title = {How much power does your server consume? Estimating wall socket power using {RAPL} measurements},
year = {2016},
issue_date = {November  2016},
publisher = {Springer-Verlag},
address = {Berlin, Heidelberg},
volume = {31},
number = {4},
issn = {1865-2034},
doi = {10.1007/s00450-016-0325-4},
journal = {ComputerScience - Research and Development},
month = nov,
pages = {207–214},
numpages = {8}
}

@online{altosafemode,
    title = {Pimlico - Self-Host Guide},
    year = {2025},
    url = {https://docs.pimlico.io/references/bundler/self-host},
    lastaccessed = {September 19, 2025}
}

@online{bundlersoverview,
    title = {{ERC}-4337 Bundlers},
    year = {2025},
    url = {https://bundle.rs/},
    lastaccessed = {September 19, 2025}
}

@online{powerapiorg,
    title = {{PowerAPI} - Measuring power consumption of your applications},
    year = {2025},
    url = {https://powerapi.org/},
    lastaccessed = {September 19, 2025}
}

@online{anvil,
    title = {Anvil - A fast local {Ethereum} development node},
    year = {2025},
    url = {https://getfoundry.sh/anvil/overview/},
    lastaccessed = {September 19, 2025}
}

@online{infinitism,
    title = {Infinitism - {A}ccount abstraction},
    year = {2025},
    url = {https://github.com/eth-infinitism/account-abstraction},
    lastaccessed = {September 19, 2025}
}

@online{entrypoint6,
    title = {{EntryPoint v0.6.0 - Etherscan}},
    year = {2025},
    url = {https://etherscan.io/address/0x5ff137d4b0fdcd49dca30c7cf57e578a026d2789},
    lastaccessed = {September 19, 2025}
}

@online{entrypoint7,
    title = {{EntryPoint v0.7.0 - Etherscan}},
    year = {2025},
    url = {https://etherscan.io/address/0x0000000071727de22e5e9d8baf0edac6f37da032},
    lastaccessed = {September 19, 2025}
}

@inproceedings{Pereira2017,
author = {Pereira, Rui and Couto, Marco and Ribeiro, Francisco and Rua, Rui and Cunha, J\'{a}come and Fernandes, Jo\~{a}o Paulo and Saraiva, Jo\~{a}o},
title = {Energy efficiency across programming languages: how do energy, time, and memory relate?},
year = {2017},
isbn = {9781450355254},
publisher = {Association for Computing Machinery},
doi = {10.1145/3136014.3136031},
booktitle = {Proceedings of the 10th ACM SIGPLAN International Conference on Software Language Engineering},
pages = {256–267},
numpages = {12},
location = {Vancouver, BC, Canada},
}

@online{altobundler,
    title = {{Alto Bundler}},
    year = {2025},
    url = {https://docs.pimlico.io/references/bundler},
    lastaccessed = {September 19, 2025}
}

@INPROCEEDINGS{Lin2024,
  author={Lin, Zibin and Wang, Taotao and Zhao, Chonghe and Zhang, Shengli and Yang, Qing and Shi, Long},
  booktitle={2024 International Conference on Computing, Networking and Communications (ICNC)}, 
  title={A Measurement Investigation of {ERC}-4337 Smart Contracts on {E}thereum Blockchain}, 
  year={2024},
  volume={},
  number={},
  pages={1164-1170}}

@INPROCEEDINGS{Wang2022,
  author={Wang, Zicheng and Cui, Bo and Hou, Wenhan},
  booktitle={2022 IEEE 46th Annual Computers, Software, and Applications Conference (COMPSAC)}, 
  title={A Dynamic Load Balancing Scheme Based on Network Sharding in Private {E}thereum Blockchain}, 
  year={2022},
  volume={},
  number={},
  pages={362-367}}
}

\end{document}